\documentclass[aps,prb,twocolumn,superscriptaddress]{revtex4-1}
\usepackage{amsmath,amssymb,mathrsfs}
\usepackage{natbib}
\usepackage{subfigure}
\usepackage{tabularx}
\usepackage{epsfig}
\usepackage{longtable}
\usepackage{amsfonts}
\usepackage{rotating}
\usepackage{bbold}
\usepackage{hhline}
\usepackage{braket}
\usepackage{txfonts, comment}

\usepackage{appendix}
\usepackage[unicode=true,bookmarks=true,bookmarksnumbered=false,bookmarksopen=false,breaklinks=false,pdfborder={0 0 1},backref=false,colorlinks=true]{hyperref}

\hypersetup{linkcolor=magenta,urlcolor=blue,citecolor=blue,pdfstartview={FitH},hyperfootnotes=false,unicode=true}

\def\be{\begin{equation}}
\def\ee{\end{equation}}
\def\bea{\begin{eqnarray}}
\def\eea{\end{eqnarray}}
\def\nn{\nonumber}

\def\im{{\rm i}}

\begin{document}

\title{Programmable Hamiltonian engineering with quadratic quantum Fourier transform}

\author{Pei Wang}
\altaffiliation{These authors contributed equally to this work.} 
\affiliation{Department of Physics, Zhejiang Normal University, Jinhua 321004, China}
\author{Zhijuan Huang} 
\altaffiliation{These authors contributed equally to this work.} 
\affiliation{Department of Physics, Zhejiang Normal University, Jinhua 321004, China}
\author{Xingze Qiu} 
\altaffiliation{These authors contributed equally to this work.} 
\affiliation{State Key Laboratory of Surface Physics, Institute of Nanoelectronics and Quantum Computing, and Department of Physics, Fudan University, Shanghai 200438, China}
\affiliation{Shanghai Qi Zhi Institute, Shanghai 200030, China}
\author{Xiaopeng Li}  
\email{xiaopeng\underline{ }li@fudan.edu.cn}
\affiliation{State Key Laboratory of Surface Physics, Institute of Nanoelectronics and Quantum Computing, and Department of Physics, Fudan University, Shanghai 200438, China}
\affiliation{Shanghai Qi Zhi Institute, Shanghai 200030, China}
\affiliation{Shanghai Research Center for Quantum Sciences, Shanghai 201315, China}

\begin{abstract}
 Quantum Fourier transform (QFT) is a widely used building block for quantum algorithms, whose scalable implementation is challenging in experiments. Here, we propose a protocol of quadratic quantum Fourier transform (QQFT), considering cold atoms confined in an optical lattice. This QQFT is equivalent to QFT in the single-particle subspace, and produces a different unitary operation in the entire Hilbert space. 
 We show this QQFT protocol can be implemented using programmable laser potential with the digital-micromirror-device techniques recently developed in experiments.  The QQFT protocol enables programmable Hamiltonian engineering, and allows quantum simulations of Hamiltonian models, which are difficult to realize with conventional approaches. The flexibility of our approach is demonstrated by performing quantum simulations of one-dimensional Poincar\'{e} crystal physics and two-dimensional topological flat bands, where the QQFT protocol effectively generates the required long-range tunnelings despite the locality of the cold atom system. We find the discrete Poincar\'{e} symmetry and topological properties in the two examples respectively have robustness against a certain degree of noise that is potentially existent in the experimental realization. We expect this approach would open up wide opportunities for optical lattice based programmable quantum simulations. 
\end{abstract}

\date{\today}

\maketitle

\section{Introduction}

Quantum Fourier transform~\cite{coppersmith1994} has been widely used in constructing efficient quantum algorithms, that have  exponential quantum speedup over the classical computing~\cite{2002_Chuang_Book,2021_Chuang_PRXQ}. 
The famous example is Shor's algorithm, where QFT is integrated in a quantum circuit to perform prime factorization of vital importance to cryptography~\cite{1994_Shor}.  
It has been combined with  control unitary circuits constituting quantum phase estimation~\cite{kitaev1995quantum}, which can compute many-body Hamiltonian spectra~\cite{2002_Chuang_Book}. 
It has also been applied to implement quantum phase kickback in digital quantum simulations of Hamiltonian time evolution~\cite{2008_Alan_PNAS}. However, its experimental realization meets much challenge~\cite{1997_Chuang_Science, 1997_Timothy_PNAS,1998_Hansen_Nature,2009_Jeremy_Science,2012_Lucero_NatPhys,2016_Blatt_Science},  with present noisy-intermediate-scale-quantum devices~\cite{2018_Preskill_Quantum}. Large scale implementation so-far has not been achieved.

Cold atoms confined in optical lattices provide a versatile platform for large-scale quantum simulations of quantum many-body physics. There has been tremendous progress in simulating strongly correlated equilibrium physics and exotic quantum dynamics in cold atom experiments. The Fermi-Hubbard model of fundamental importance to modeling correlated electrons has been implemented with upto hundreds of atoms~\cite{2002_Zoller_PRL,2013_Greif_Uehlinger_Science,mazurenko2017cold,2015_Hart_Duarte_Nature,2019_Bakr_BadMetal}. Mott-superfluid transition and its quantum criticality  have been studied in lattices of different geometry and of different dimensionality~\cite{2015_Lewenstein_Review,2016_Li_RPP,2017_Gross_Science}. Topological bands and geometrical Berry phases in momentum space~\cite{2013_Bloch_NatPhys,2014_Esslinger_Nature,2015_Goldman_NatPhys,2015_Ketterle_NatPhys,2016_Pan_Science,2016_Sengstock_Science,2021_Pan_Science} have been observed in artificial gauge field lattices with the effective magnetic field  strength reaching the order of unit flux quantum per unit cell. The dynamical phase transition from quantum thermalization to many-body-localization have been investigated with both interaction and disorder strengths under control~\cite{kohlert2019observation,lukin2019probing}. 
 In these fascinating accomplishments, a broad range of local Hamiltonian models have been engineered in cold atom experiments, which has been a driving force in studying quantum many-body physics in extreme regimes in the last decade.

In more recent years, the spatial-resolved control of atom-confining potential has been achieved using digital-micromirror-device (DMD)~\cite{ha2015roton, gauthier2016direct, 2016_Weiss_Science, mazurenko2017cold, browaeys2020many} or sub-wavelength techniques~\cite{Yi_2008,2019_Zoller_PRA,2021_Sengstock_Nature} in the experiments, which  
does not only  allow high resolution imaging of the atomic system, but also enables programmable Hamiltonian engineering with local potential energies on each individual lattice site  being tunable  one by one~\cite{2016_Weiss_Science,browaeys2020many,2021_Sengstock_Nature}, with even the simulation error being correctable~\cite{Qiu2020npjQI,lukin2019probing}. 
These single-site techniques provide  unprecedented opportunities for cold atom quantum simulations, for example in probing localization physics with programmable disorders~\cite{Qiu2020npjQI}, realizing large scale quantum spin glasses, and solving binary optimization problems~\cite{2015_Zoller_SA,2020_Qiu_PRXQ}.

Here, we construct a scheme of  quadratic quantum Fourier transform, which is equivalent to the full version QFT in the single-particle subspace, and show how 
to implement this scheme in cold atom experiments. 
This QQFT scheme gives rise to highly flexible Hamiltonian engineering. The momentum space properties of the engineered Hamiltonian model including both band structure and Berry curvature become completely programmable with the QQFT scheme, which is beyond the capability of conventional Floquet engineering approach~\cite{2017_Eckardt_RMP,2019_Cooper_RMP}.  
This scheme is applied to quantum simulations of $1+1$D Poincar\'{e} crystals and 2D topological flat bands. Despite the requirement of long-range tunnelings in the direct implementation of these models, they become accessible to optical lattice experiments with our QQFT scheme. We find the QQFT based Hamiltonian engineering protocols have reasonable robustness against experimental imperfections.
The QQFT scheme with the ongoing experimental developments on single-site control is expected to add a novel dimension for optical lattice based quantum simulations. 

This paper is organized as follows. In Sec.~\ref{sec:modmeth}, we introduce
the idea of Hamiltonian engineering with QQFT, show how to construct QQFT by using
local operations that are accessible in the experiments, and
explain the difference of QQFT from QFT. In Sec.~\ref{sec:poin} and Sec.~\ref{sec:chern},
we apply the QQFT scheme to simulations of the 1D Poincar\'{e} crystal and flat-band Haldane
model, respectively, and discuss the robustness of their properties against noise.
Section~\ref{sec:con} is a short summary.

\section{Model and method}
\label{sec:modmeth}

\begin{figure}[htp]
\centering
\includegraphics[width=.4\textwidth]{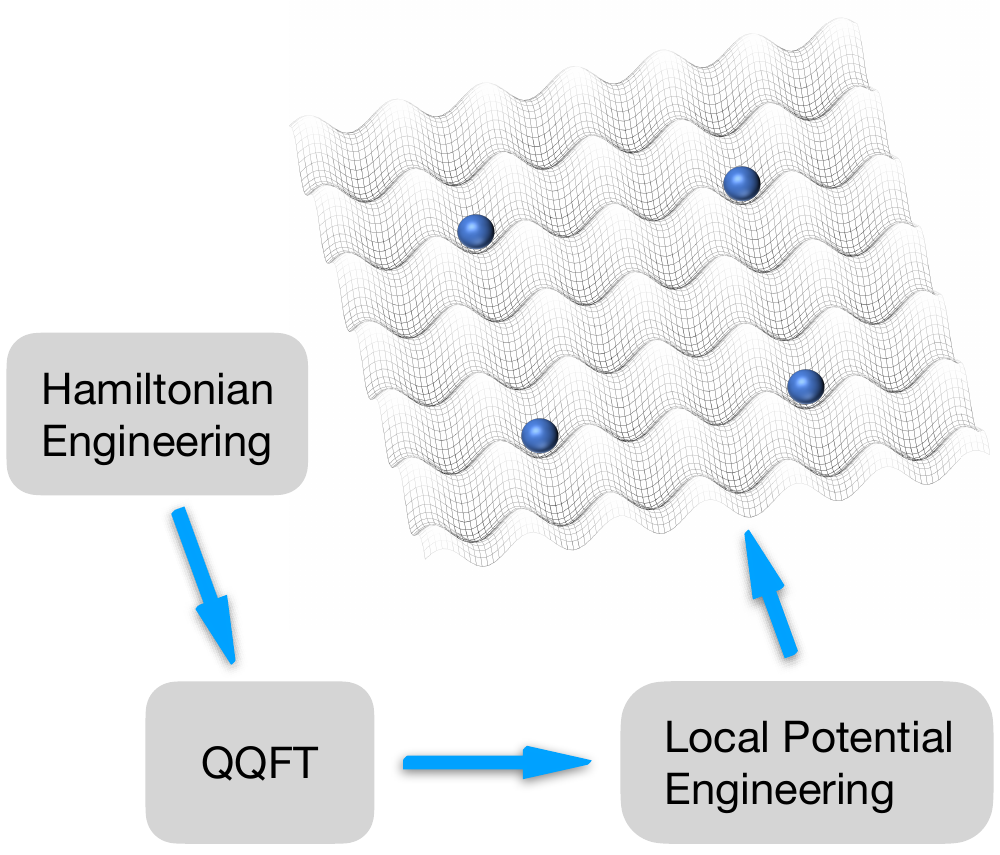}
\caption{Schematic illustrations of the Hamiltonian engineering protocol
through QQFT, which can be implemented with local potential engineering in the optical lattice platform.
}
\label{fig:OL}
\end{figure}

\subsection{One-dimensional single-band model}

We first take a one-dimensional single-band model to illustrate  the main idea of
Hamiltonian engineering with QQFT for simplicity.
Let us assume that the Hamiltonian ($\hat{H}$) model we aim at simulating in the experiment 
has a lattice translation symmetry, so it has a form 
$\hat{H}= \sum_{m,n} 
J_{m-n} \hat{\psi}^\dag_{m}
\hat{\psi}_{n}$, where $J_{m-n}$ is the hopping amplitude, 
and $\hat{\psi}_{n}$ ($\hat{\psi}^\dag _m $) is the annihilation (creation) operator at the lattice site $n$.

Although a large range of Hamiltonian models have been achieved in cold atom experiments, engineering long-range coupled Hamiltonians is extremely challenging with this system for its locality and diluteness~\cite{2017_Gross_Science}. Now, we develop a protocol to simulate the Hamiltonian quantum dynamics $\hat{U} =e^{-i\hat{H}T}$, which applies to all translationally invariant models,  regardless of whether long-range couplings exist or not. 

The idea is that the evolution operator $\hat{U}$ takes a diagonal form by a Fourier transform, 
$\hat{U} = \hat{V} e^{-i\hat{H}_{\rm D} T} \hat{V}^\dag$, 
with 
\bea 
\label{eq:orHam}
&& \hat{H}_{\rm D}=\sum_{m} 
\hat{\psi}^\dag_{m }
\mathcal{H} \left(  m \right) 
\hat{\psi}_m,   \\ 
&& 
\hat{V} = \exp \left( \frac{2\pi i}{L} \sum_{  n,  m }  
A_{m, n}
\hat{ \psi} _{m} ^\dag   
\hat{ \psi} _{n} \right), 
\label{eq:Fourier} 
\eea 
where $A$ is a Hermitian matrix defined by
$\left(\omega ^{A}\right)_{m,n}
= \frac{1}{\sqrt{L}} \omega^{mn}$ 
with 
$\omega=e^{i2\pi/L}$, 
and $L$ the number of lattice sites. Here we have used
\begin{equation}\label{eq:omegaquaiden}
\omega^{\hat{\Psi}^\dag A_1 \hat{\Psi}} \omega^{\hat{\Psi}^\dag A_2 \hat{\Psi}} 
= \omega^{\hat{\Psi}^\dag A_3 \hat{\Psi}}
\end{equation}
with $\omega^{A_1} \omega^{A_2}=\omega^{A_3}$ and $\hat{\Psi}=\left(\hat{\psi}_1,\hat{\psi}_2,\cdots,\hat{\psi}_L \right)^T$.
The array of $\mathcal{H}(m)$ contains  the Fourier transform of the tunneling matrix of $J_{m-n}$, i.e.,
\be 
{\cal H} (m) =  \sum_{n'}  \omega ^{-mn'} J_{n'}  . 
\ee
The quantum dynamics by $\hat{U}$ then involves the local phase evolution ($e^{-i\hat{H}_{\rm D} T}$), and  the QQFT ($\hat{V}$). 
By QQFT, engineering the momentum space Hamiltonian $\mathcal{H} ( m )$, or equivalently the band structure and the momentum space Berry phase~\cite{2017_Eckardt_RMP}, is reduced to programming the local potential, which is accessible with present optical lattice techniques~\cite{2016_Weiss_Science,browaeys2020many,2021_Sengstock_Nature,Qiu2020npjQI} (see Fig.~\ref{fig:OL} for a schematic illustration).

The key is how to implement QQFT in the experimental system, whose
direct realization involves highly nontrivial long-range couplings. 
With the numerical  decomposition scheme for generic unitary operations in Ref.~[\onlinecite{Qiu2020npjQI}], the QQFT can be obtained by a sequential Hamiltonian evolution of a depth, $D \propto L^2$. 
We further exploit the mathematical structure of Fourier transform, and construct an analytic Hamiltonian sequence $\hat{H}_{\rm p} ^{[s=1,2,\ldots D]}$
such that 
\be 
\label{eq:HamSeq} 
\textstyle 
\hat{V}  = 
e^{-i\hat{H}_{\rm p} ^{[D]} } 
\times 
e^{-i\hat{H}_{\rm p} ^{[D-1]}} 
\times 
\ldots 
\times 
e^{-i\hat{H}_{\rm p} ^{[1]}}, 
\ee 
where all the Hamiltonians are strictly local involving local potential and nearest neighboring tunneling only. 
For a system size being an integer power of $2$, we have a Hamiltonian sequence depth, 
\be \label{eq:expDD}
\textstyle D = L (\log_2 L +2)/2 -\log_2 L  -1,  
\ee 
which scales as $D\propto L\log L$ for large system size. The detail will be
explained in next section.

\subsection{Local construction of 1D QQFT} 
\label{sec:subsecQQFT}

In this section, we give the analytic construction of QQFT using local gate operations.
The one dimensional QQFT (Eq.~\eqref{eq:Fourier}) can be reexpressed as
\be 
\label{eq:V1D}
\hat{V} =
\omega ^{\hat{\Psi}^\dag A \hat{\Psi} } .
\ee 
For such operators $\omega^{\hat{\Psi}^\dag A \hat{\Psi}}$, Eq.~\eqref{eq:omegaquaiden} stands,
which implies that a quadratic Hamiltonian sequence producing the QQFT in the single-particle subspace necessarily gives the QQFT unitary operator in Eq.~\eqref{eq:V1D}. In the following, we present the construction of QQFT using single-particle basis 
to save writing. 

\begin{figure*}[htp]
\centering
\includegraphics[width=.7\textwidth]{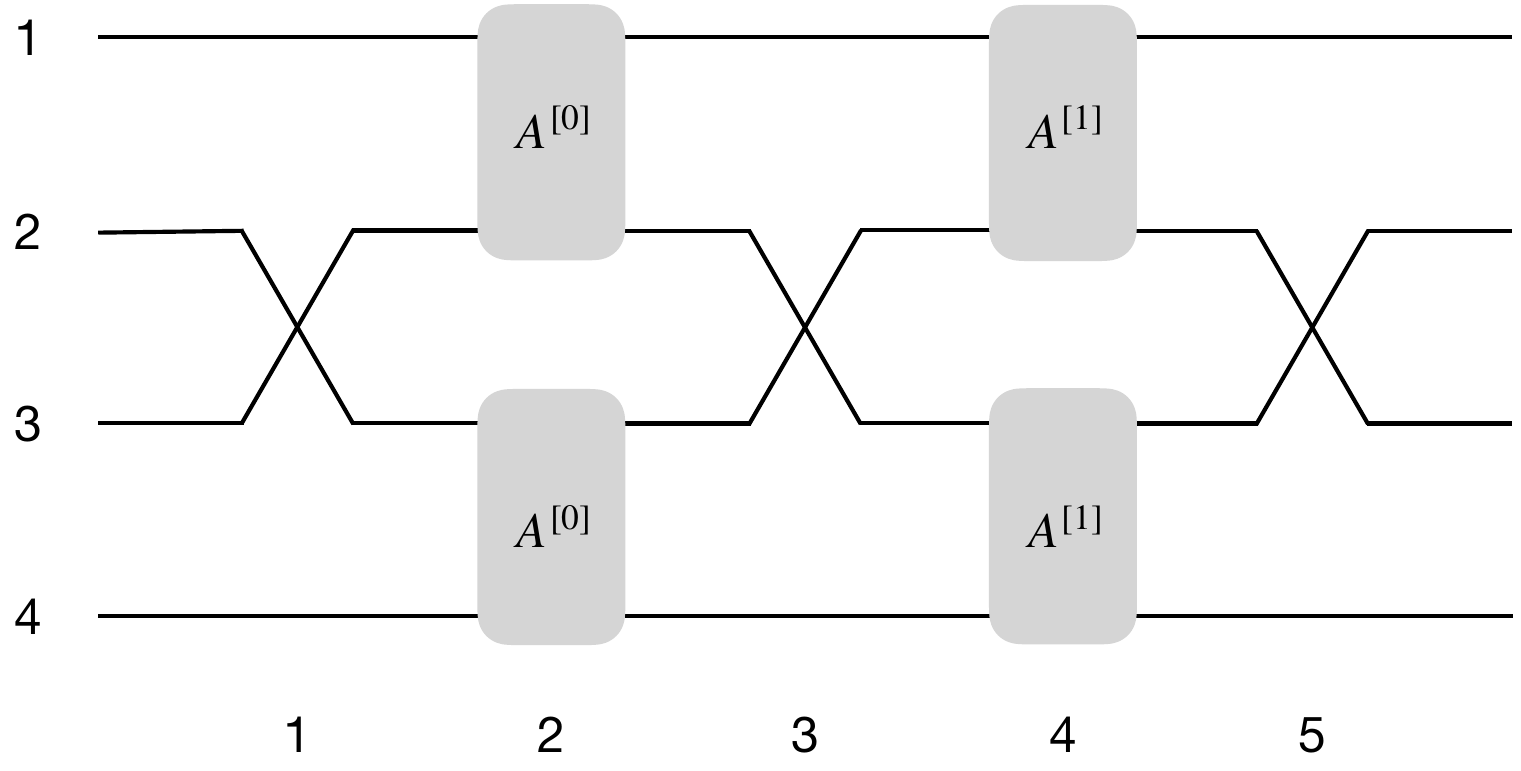}
\caption{Schematic illustrations of the construction of QQFT using local gate operations. 
The crosses and the hoary rounded rectangles indicate the swap operations ${\cal R}$ and the local unitaries ${\cal A}$, respectively. 
Here, the input state is fed into the circuit from left, and we choose $L=4$ for illustration.
}
\label{fig:QQFT}
\end{figure*}

We assume the number of lattice sites is an integer power of $2$, i.e., $L = 2^l$. 
Introducing a binary representation of $k$ and $j$, as 
\bea 
k &=& (k_{l-1}, k_{l-2}, \ldots, k_0) \nn \\
j &=& (j_{l-1}, j_{l-2}, \ldots, j_0) ,   \nn 
\eea 
the Fourier transform has a sequential product form 
\be 
\label{eq:omegaproduct} 
\omega^{kj} = 
\omega ^{2^{l-1} k_0 j_{l-1}} 
\omega ^{(2^{l-1} k_1 + 2^{l-2} k_0) j_{l-2}} 
\ldots 
\omega ^{(2^{l-1} k_{l-1} + 2^{l-2} k_{l-2} +\ldots + 2^0 k_0) j_0} .
\ee 
This product form has been used to construct the fast Fourier transform algorithm in classical computing~\cite{1988_Brigham_FFT}. 
For our quantum gate construction, the product form implies the matrix $\Omega$, defined by 
$\Omega_{kj} = \frac{1}{\sqrt{L}} \omega^{kj}$, 
can be rewritten as a sequence of unitary matrix operations, 
\be 
\Omega  = \Omega ^{[l-1]} \Omega^{[l-2]} \ldots \Omega^{[1]} \Omega^{[0]}, 
\ee 
with the matrix $\Omega^{[q]}$ defined by 
\be 
\begin{split}
\Omega^{[q]}_{kj}   = & \frac{1}{\sqrt{2}} \omega ^{(2^{l-1} k_{q} + 2^{l-2} k_{q-1} + \ldots + 2^{l-q-1 } k_0 ) j_{l-1}} 
\delta_{k_{l-1}, j_{l-2}} \delta_{k_{l-2}, j_{l-3}} \ldots \\ & \times
\delta_{k_{q+1}, j_{q}}  
\delta_{k_{q-1}, j_{q-1}} 
\delta_{k_{q-2}, j_{q-2}} \ldots 
\delta_{k_{0}, j_{0}}. 
\end{split}
\ee 

We now construct $\Omega^{[q]}$ using local gate operations. 
To proceed, we define a re-ordering unitary ${\cal R}^{[p]} $ as  
\be 
{\cal R}^{[p]}_{kj} = 
\delta_{k_{l-1}, j_{l-1}} 
\delta_{k_{l-2}, j_{l-2}} 
\ldots 
\delta_{k_{p+1}, j_{p+1} }
\delta_{k_{p}, j_{0}}
\delta_{k_{p-1}, j_{p} }
\delta_{k_{p-2}, j_{p-1}} 
\ldots
\delta_{k_0, j_1 }. 
\ee 
This re-ordering unitary can be implemented by a series of swap operations between  neighboring sites, which are accessible to programmable optical lattice experiments~\cite{Qiu2020npjQI}. 
The neighboring swap operations between $j$ and $j'$ are denoted as $(j, j')$. 
We introduce a composite swap operation,  
\be 
{\cal S}^{[jj']} = (j,j+1) (j+2, j+3) \ldots (j'-1, j'), 
\ee 
which involves multiple swap operations being parallelizable. 
The ${\cal R}^{[p]}$-unitary is then given by 
\be 
\begin{split}
{\cal R} ^{[p]} 
= & \prod_{\gamma=0} ^ {2^{l-p-1}-1 } 
    {\cal S}^{[2^{p+1}\gamma  +2^{p}-1, 2^{p+1}\gamma + 2^{p}] }
    {\cal S}^{[2^{p+1}\gamma+2^{p}-2, 2^{p+1}\gamma+ 2^{p}+1] } 
    \ldots \\ & \times
    {\cal S}^{[2^{p+1}\gamma+2, 2^{p+1}\gamma+2^{p+1}-3]} 
    {\cal S}^{[2^{p+1}\gamma+1, 2^{p+1}\gamma+2^{p+1}-2 ]}.  
\end{split}
\ee 
In terms of the swap operations, the unitary matrix $\Omega^{[q]}$ is rewritten as 
\be 
\Omega ^{[q]} = {\cal R} ^{[q]}  {\cal A}^{[q]} {\cal R}^{[l-1]\dag },  
\ee 
with ${\cal A}$ a local unitary, 
\be 
{\cal A}^{[q]} _{k j}
= 
\delta_{k_{l-1}, j_{l-1}}  
\delta_{k_{l-2}, j_{l-2}} 
\ldots 
\delta_{k_{1}, j_{1}} 
A^{[q]}_{k_0, j_0}, 
\ee 
and $A^{[q]}$ a matrix 
\be 
A^{[q]} 
= \frac{1}{\sqrt{2}} 
\left[ 
\begin{array}{cc} 
 1 &  \omega^{(2^{l-2} j_{q} +2^{l-3} j_{q-1} + \ldots+ 2^{l-q-1}j_1  )}  \\  
 1 & -\omega^{(2^{l-2} j_{q} +2^{l-3} j_{q-1} + \ldots+ 2^{l-q-1}j_1  )}   
\end{array} 
\right]. 
\ee 

The number of local gate operations constituting $\Omega^{[q]}$ is $2^{l-1}+2^{q} -1 $. The total number of local gate operations for the whole 1D QQFT is then 
\be 
D  = \sum_{q=0}^{l-1} 
\left[  2^{l-1}+2^{q} -1  \right] 
=  \left(2^{l-1}-1\right) l + 2^l -1,
\ee 
which is equivalent to Eq.~\eqref{eq:expDD}.
The sequential unitary operations, which contain the sequential swap operations ${\cal S}^{[jj']}$ and the local unitaries $A^{[q]}$, 
are then realized by a Hamiltonian sequence $\hat{H}_{\rm p}^{[s=1,2,\ldots D]}$ with depth $D$. 
One explicit example with $L= 4$ can be found in Fig.~\ref{fig:QQFT}. In this case, the  Hamiltonian sequence ($\hat{H}_p^{s}$)  for realizing the QQFT has a depth $D = 5$, with 
\be
\begin{split}
\hat{H}_p^{[1]} =&\hat{H}_p^{[3]} = \hat{H}_p^{[5]} =
    \frac{\pi}{2} \left[  \hat{\psi}_2 ^\dag \hat{\psi}_3 + \hat{\psi}_3 ^\dag \hat{\psi}_2
    - \hat{\psi}_2 ^\dag \hat{\psi}_2-\hat{\psi}_3 ^\dag \hat{\psi}_3 \right]  \\ 
\hat{H}_p^{[2]} =& \frac{\pi}{2} 
\left[ \hat{\psi}_1 ^\dag \hat{\psi}_2 + \hat{\psi}_2^\dag \hat{\psi}_1 
+ \hat{\psi}_3^\dag \hat{\psi}_4 + \hat{\psi}_4 ^\dag \hat{\psi}_3 
    + \hat{\psi}_1^\dag \hat{\psi}_1 + \hat{\psi}_3 ^\dag \hat{\psi}_3 
    - \hat{\psi}_2 ^\dag \hat{\psi}_2 \right. \\ & \left. -\hat{\psi}_4 ^\dag \hat{\psi}_4 \right]/\sqrt{2} 
   - \frac{\pi}{2} \left[ \hat{\psi}_1 ^\dag \hat{\psi}_1 + \hat{\psi}_2 ^\dag \hat{\psi}_2
       +  \hat{\psi}_3 ^\dag \hat{\psi}_3 + \hat{\psi}_4 ^\dag \hat{\psi}_4 \right] \\ 
\hat{H}_p^{[4]} =& \frac{\pi}{2} 
\left[ (\hat{\psi}_1 ^\dag \hat{\psi}_2 + \hat{\psi}_2^\dag \hat{\psi}_1  
    + \hat{\psi}_1^\dag \hat{\psi}_1 - \hat{\psi}_2 ^\dag \hat{\psi}_2)/\sqrt{2}  \right. \\ & \left. 
    + (\hat{\psi}_3 ^\dag, \hat{\psi}_4 ^\dag ) {O} (\hat{\psi}_3, \hat{\psi}_4 ) ^T
    \right]- \frac{\pi}{2} 
\left[ \hat{\psi}_1 ^\dag \hat{\psi}_1 + \hat{\psi}_2 ^\dag \hat{\psi}_2 \right] ,
\end{split}
\ee  
with $O$ a $2\times 2$ matrix defined by 
\be 
e^{-i\pi O/2} = \frac{1}{\sqrt{2}} 
\left[ 
\begin{array}{cc}
1 & i \\ 
1 & -i 
\end{array}
\right ]. 
\ee 
The elementary unitary transformation between two neighboring sites is always accessible to the programmable optical lattices without synthetic gauge fields~\cite{Qiu2020npjQI}.   
The code producing the Hamiltonian sequence for arbitrary $L$ is available upon reasonable request. 

The above construction relies on the assumption that $L$ is an integer power of $2$. 
For a generic $L$ that does not take this form, we shall factorize $L$ into a product of prime numbers.  
The QQFT can be decomposed into a series of QQFT blocks corresponding to the factorization, drawing an analogy to  Eq.~\eqref{eq:omegaproduct}. Each prime QQFT block can be constructed following the scheme provided in Ref.~[\onlinecite{Qiu2020npjQI}]. 

Considering Li atoms~\cite{Hulet2020Review} confined in an optical lattice
with a laser wavelength $1064$ nm, whose recoil energy is $E_{\rm R}/\hbar = 2\pi\times25.12$kHz, 
one cycle of 1D QQFT takes about $100$ ms for $L = 32$, using a tunneling strength $J= 0.01 E_{\rm R}$. 
This required time scale is accessible to the present cold atom experiments. 

With the QQFT based quantum simulation, any Hamiltonian model that has
translational symmetry and is consequently diagonal in momentum space,
can be realized by programming the local optical lattice potential. This would
give unprecedented opportunities for cold atom based quantum simulations. 
Moreover, the QQFT approach provides a generic recipe to engineer long-range interacting models for local quantum computing architectures such as Rydberg atoms~\cite{2010_Saffman_RMP} and superconducting qubits~\cite{2020_Kjaergaard_Review}. 
The implementation cost with the $L{\rm log} L$ scaling is nearly optimal---a lower bound linear to $L$ is expected according to the Lieb Robinson bound~\cite{1972_Lieb}. 

\subsection{Distinction between QQFT and standard QFT}

{This proposing QQFT is only equivalent to the full QFT in the single-particle Hilbert space. It produces a different transformation in the entire Hilbert space.}
Here we discuss about the connection between QQFT (denoted as $\hat{V}$)
and the standard QFT (denoted as $\hat{F}$), and clarify their distinction. 

One drastic difference between QQFT and standard QFT is about particle number conservation. 
The QQFT is defined in terms of annihilation ($\hat{\psi}$) and creation ($\hat{\psi}^\dag$) operators. The particle number conservation is explicit in QQFT. 
The standard QFT is defined using the computation basis $|z_1, z_2, \ldots, z_L\rangle$ ($z_j = 0,1$), where the number of particles can be defined as $\sum_j z_j$. A standard basis index $x$ is defined by taking $(z_1, z_2, \ldots, z_L)$ as a binary number. The matrix elements of standard QFT are
\be 
\langle x | F | x'\rangle = \frac{1}{\sqrt{2^L} }  e^{2\pi i xx' /2^L  }  .
\label{eq:sQFT}
\ee 
The standard QFT does not respect particle number conservation.

If we restrict to the single-particle subspace defined by $\sum_j z_j = 1$, the dimension of this subspace is $L$, and the basis states are then labeled by $|j\rangle$ according to the position of the particle. In this subspace, a natural way to define the standard QFT is to assume $\langle j | \hat{F} |j'\rangle = e^{2\pi ijj'/L} /\sqrt{L}$. 
In this subspace,  the matrix elements of QQFT are $\langle j | \hat{V} |j'\rangle = e^{2\pi ijj'/L} /\sqrt{L}$, which are identical to the standard QFT.

Considering multi-particle Hilbert space, the QQFT and standard QFT are completely different, and there is no meaningful way to generalize standard QFT to make it identical to QQFT. Besides the aspect of particle number conservation, the other key difference is about quantum statistics. Since the QQFT is defined in terms of annihilation  and creation   operators, the precise definition of QQFT actually depends on whether the $\hat{\psi}$ operators are bosonic or fermionic. With bosonic operators, the local occupation number could be $0$ upto the total particle number, which makes the corresponding bosonic QQFT completely different from the standard QFT. With fermionic operators, the QQFT is also different from standard QFT. To see their difference, we take the example of two-particle state, $|m_1 m_2 \rangle = \hat{\psi}^\dag_{m_1} \hat{\psi}^\dag_{m_2} |{\rm vaccum} \rangle$. In the two-particle subspace, the matrix elements of QQFT are 
\be 
\langle m_1 m_2 | \hat{V} |n_1 n_2 \rangle 
= \frac{1}{L} \left[ e^{2\pi i (m_1 n_1 + m_2 n_2)/L} - e^{2\pi i (m_1 n_2 + m_2 n_1)/L}  \right]. 
\ee 
This is completely different from the standard QFT (Eq.~\eqref{eq:sQFT}), and there is no meaningful way to generalize the standard QFT to make it identical to the QQFT.

\subsection{General framework of Hamiltonian engineering with QQFT}
\label{sec:genqqft}
 
In this section, we show how to simulate a generic $k$-band model
in a $d$-dimensional space for arbitrary $k$ and $d$. The model Hamiltonian
has a translation symmetry and is generally expressed as
\bea
\hat{H}=\displaystyle\sum_{{\bf n \alpha},{\bf n' \beta}} 
J_{{\alpha},{\beta}}\left(\bf n-\bf n'\right) \hat{\psi}^\dag_{\alpha, {\bf r}_{\bf n}}
\hat{\psi}_{\beta, {\bf r}_{{\bf n}'} }, 
\eea
where $J_{{\alpha},{\beta}}\left(\bf n-\bf n'\right)$ is the hopping amplitude, 
and $\hat{\psi}_{\alpha, {\bf r}_{\bf n} }$ is the field operator at the position 
${\bf r}_{\bf n} =\sum^d_{i=1} {n_i}{\bf a}_i$, with the orbital index $\alpha=1,2\ldots k$.
Here ${\bf a}_1$, ${\bf a}_2,\cdots$, ${\bf a}_d$ denote the primitive vectors of a Bravais lattice
in a $d$-dimensional space. Since our scheme is to emulate the momentum-space of the targeting model with the real-space lattice of a physical system, it is natural to assume a simple cubic geometry for the physical lattice, which corresponds to setting a cubic grid for the momentum-space Brillouin zone. This does not compromise the generality of the Hamiltonian engineering protocol. We assume the lattice size is $L$ in the direction of each ${\bf a}_i$, and then the total number of sites is $L^d$.

Just as in the one-dimensional single-band case, we employ the Fourier transform and factorize the evolution operator ($\hat{U}$) into $\hat{V} e^{-i\hat{H}_{\rm D} T} \hat{V}^\dag$, with 
\bea 
&& \hat{H}_{\rm D}=\sum_{\bf m, \alpha,\beta} 
\hat{\psi}^\dag_{\alpha, {\bf r}_{\bf m}}
\mathcal{H}_{\alpha \beta} \left( {\bf m} \right) 
\hat{\psi}_{\beta, {\bf r}_{\bf m}},   \\ 
&& 
\hat{V} = \exp \left( \frac{2\pi i}{L} \sum_{\alpha, {\bf n}, {\bf m}}  
A_{{\bf n},{\bf m}}
\hat{ \psi} _{\alpha, {\bf r}_{\bf n}} ^\dag   
\hat{ \psi} _{\alpha, {\bf r}_{\bf m}}\right), 
\label{eq:Fourier2} 
\eea 
where $A$ is a Hermitian matrix defined by
$\left(\omega ^{A}\right)_{{\bf n},{\bf m}}
= \omega^{{\bf n}\cdot {\bf m}}/\sqrt{L^d}$ with $\omega=e^{i2\pi/L}$. 
The matrix $\mathcal{H}({\bf m} )$ is the Fourier transform of the tunneling matrix, i.e., $\mathcal{H}_{\alpha\beta}({\bf m} ) = \displaystyle\sum_{\bf n} J_{{\alpha},{\beta}}\left(\bf n \right) e^{-i 2\pi \bf n \cdot \bf m}$. 

The phase evolution ($e^{-i\hat{H}_{\rm D} T}$) is local, and then can be simulated easily. 
More importantly, the QQFT in a $d$-dimensional space can be factorized into a sequence of one-dimensional QQFTs, which are along the directions of ${\bf a}_i$ with $i=1,2,\cdots,d$, respectively. To see it, we notice $\omega^{{\bf n}\cdot {\bf m}} = \displaystyle\prod_{i=1}^d \omega^{{n_i}{m_i}}$, therefore, the matrix $\omega^A$ is the Kronecker product of $d$ same matrices, which reads
\bea\label{eq:omegaAKro}
\begin{split}
\omega^A = & \ \omega^{\bar A} \otimes \omega^{\bar A} \otimes \cdots \otimes \omega^{\bar A} \\  = & \ \left(\omega^{\bar A} \otimes 1 \otimes \cdots \otimes 1 \right)\left(1\otimes \omega^{\bar A} \otimes \cdots \otimes 1 \right) \cdots \\
& \times \left(1\otimes 1 \otimes \cdots \otimes \omega^{\bar A}\right) ,
\end{split}
\eea
where $\left(\omega^{\bar A}\right)_{n,m} = \omega^{nm}/\sqrt{L}$ is the one-dimensional QQFT matrix and $1$ is the identity matrix. The second line of Eq.~\eqref{eq:omegaAKro} tells us that $\omega^A$ is a product of $d$ matrices, i.e., $\omega^A = \displaystyle\prod_{i=1}^d \omega^{{A}_i}$, where $\omega^{{A}_i}$ is the Kronecker product of $\omega^{\bar A}$and $\left(d-1\right)$ identity matrices. According to Eq.~\eqref{eq:omegaquaiden},
we immediately know that the factorization of $\omega^A$ indicates a factorization of $\hat V$, i.e., $\hat V= \displaystyle\prod_{i=1}^d \hat V^{(i)}$ with $\hat V^{(i)}= \omega^{\hat{\Psi}^\dag A_i \hat{\Psi}} $.
Therefore, the $d$-dimensional QQFT ($\hat V$) can be realized by successively performing $\hat{V}^{(1)}$, $\hat{V}^{(2)}$, $\cdots$, $\hat{V}^{(d)}$.

According to the definition of $\omega^{A_i}$, we see that the corresponding operator $\hat V^{(i)}$ involves only the hopping in the direction of ${\bf a}_i$ but not in the other $\left(d-1\right)$ directions. Therefore, $\hat V^{(i)}$ on a $d$-dimensional cubic lattice is equivalent to the QQFT on a one-dimensional lattice in the ${\bf a}_i$-direction, which can be effectively constructed by using the numerical decomposition scheme for generic unitary operations~\cite{Qiu2020npjQI}, or by using the analytic sequence introduced in Sec.~\ref{sec:subsecQQFT}. Since the Fourier transform across different spatial dimensions is separable, the Hamiltonian sequence depth in a $d$-dimensional QQFT is exactly $d$ times of that in a one-dimensional QQFT.
A $d$-dimensional QQFT involves $d$-cyles of 1D QQFT, taking a few hundred ms,
if we consider Li atoms in an optical lattice of size $L = 32$
with a laser wavelength $1064$ nm and recoil energy $E_{\rm R}/\hbar=2\pi\times25.12$kHz,
using a tunneling strength $J= 0.01 E_{\rm R}$. This time scale is
accessible to the cold atom experiments.

\begin{widetext}
\begin{center}
\begin{table}[b]
\renewcommand\arraystretch{2.0}
\begin{tabular}{| p{1.3 cm}<{\centering} | p{0.3 cm}<{\centering} | p{0.3 cm}<{\centering}
| p{0.3 cm}<{\centering} | p{0.3 cm}<{\centering} | p{0.3 cm}<{\centering} |
p{0.3 cm}<{\centering} | p{0.3 cm}<{\centering} | p{0.3 cm}<{\centering} | p{0.3 cm}<{\centering} | p{0.3 cm}<{\centering} |  p{0.3 cm}<{\centering} |  p{0.3 cm}<{\centering} |  p{0.3 cm}<{\centering} |  p{0.3 cm}<{\centering} |  p{0.3 cm}<{\centering} |  p{0.3 cm}<{\centering} |  p{0.3 cm}<{\centering} |  p{0.3 cm}<{\centering} |  p{0.3 cm}<{\centering} |  p{0.3 cm}<{\centering} |  p{0.3 cm}<{\centering} |  p{0.3 cm}<{\centering} |  p{0.3 cm}<{\centering} |  p{0.3 cm}<{\centering} |  p{0.3 cm}<{\centering} |  p{0.3 cm}<{\centering} |  p{0.3 cm}<{\centering} |  p{0.3 cm}<{\centering} |  p{0.3 cm}<{\centering} |  p{0.3 cm}<{\centering} |  p{0.3 cm}<{\centering} |  p{0.3 cm}<{\centering} |  p{0.3 cm}<{\centering} |}
\hline
$\displaystyle\frac{k}{2\pi/\left(La\right)}$ & 0 & 1 & 2 & 3 & 4 & 5 & 6 & 7 & 8 & 9 & 10 & 11 & 12 & 13 & 14 & 15 & 16 & 17 & 18 & 19 & 20 & 21 & 22 & 23 & 24 & 25 & 26 & 27 & 28 & 29 & 30 & 31 & 32\\
\hline
$\displaystyle\frac{E_k}{2\pi/\left(LT\right)}$ & 0 & 6 & 12 & 15 & 24 & 30 & 30 & 9 & 15 & 12 & 27 & 0 & 27 & 12 & 18 & 9 & 30 & 3 & 24 & 15 & 21 & 6 & 0 & 6 & 21 & 18 & 24 & 3 & 3 & 9 & 18 & 21 & 27\\
\hline
\end{tabular}
\caption{A Lorentz-invariant dispersion relation.}\label{tab:Ek}
\end{table}
\end{center}
\end{widetext}

\section{Quantum simulation of Poincar\'{e} crystal}
\label{sec:poin}

\begin{figure*}[htp]
\centering
\includegraphics[width=.8\textwidth]{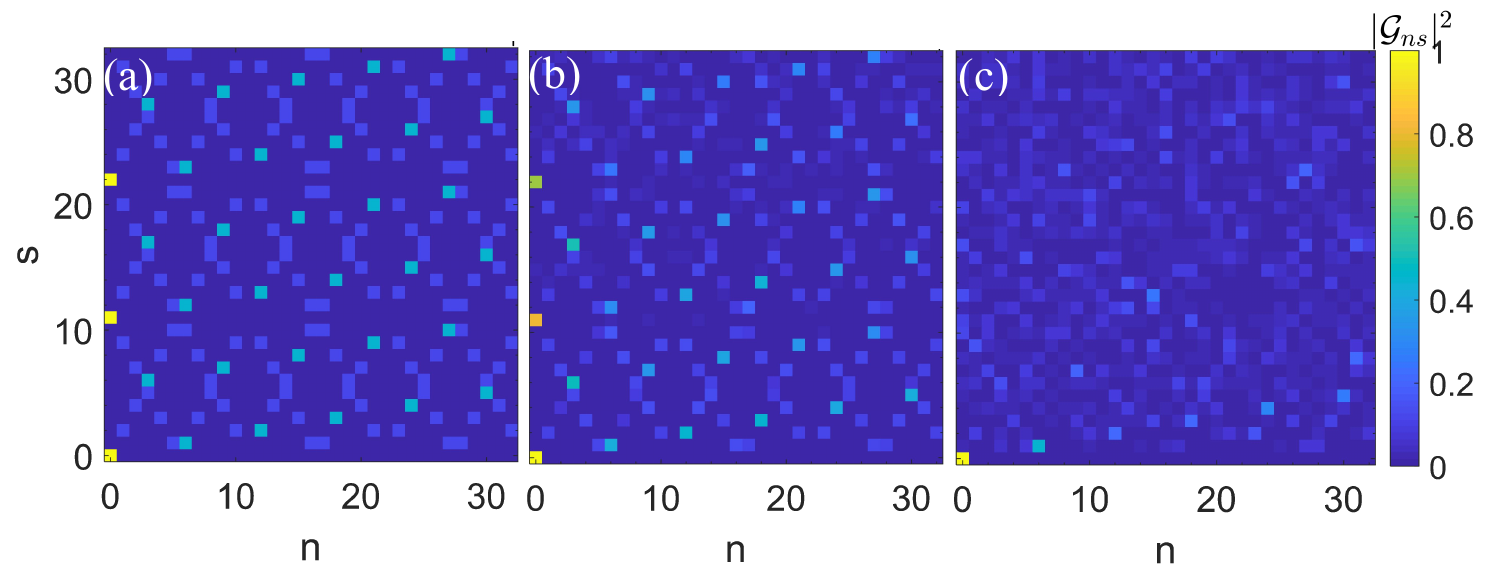}
\caption{QQFT based quantum engineering of Poincar\'{e} crystal. We simulate quantum walk of a single particle on the Poincar\'{e} crystal. The particle is initialized at $n=0$.  The color indexes $| {\cal G} _{ns}|^{2}$, the probability of finding the particle at the lattice site $n$ at the time $sT$.   Here, we consider different strength of the Gaussian white noise: (a) $\sigma=0$, (b) $\sigma=5\times10^{-3}$, and (c) $\sigma=2\times10^{-2}$. 
The system size is chosen to be $L=33$.} 
\label{fig:PC}
\end{figure*}

\subsection{Hamiltonian engineering of Poincar\'{e} crystal}

In order to benchmark our QQFT-based quantum simulation protocol, we first apply this scheme to quantum engineering of Poincar\'{e} crystal,  
which is defined by the presence of discrete spacetime translation and Lorentz symmetries~\cite{Wang2018NJP, Wang2021JPA}. 
This quantum spacetime crystal has a robust collapse and revival dynamics, i.e., crystallization in spacetime, as opposed to the typical ballistic expansion for the standard quantum wave evolution. 
Its single-particle propagator forms a Lorentz-invariant spacetime pattern.
The crystallization in spacetime is related to the discrete Lorentz symmetry, unlike the Floquet time crystal whose robustness relies on the localization and the topological $\pi$-mode~\cite{2011_Jiang_FloquetMajorana,2012_Wilczek_TC,2012_Wilczek_QTC,2016_Sondhi_PRL,2016_Nayak_PRL}. 
However, the direct experimental realization of Poincar\'{e} crystal is extremely challenging for its requirement on delicate long-range couplings~\cite{Wang2021JPA}. 

In detail, the Poincar\'{e} crystal is a lattice model whose
symmetry group is the discrete Poincar\'{e} group, which consists
of discrete spatial translations, temporal translations and Lorentz transformations.
The ratio of spatial to temporal translation periods is
$a/T=c/\sqrt{\gamma^2-1}$, where $\gamma\geq 2$ is an integer and $c$
is the invariant speed in the Lorentz transformation (e.g. the speed of light).
$\gamma$ and $c$ determine an elementary Lorentz
transformation, which in the {1+1}-dimensional spacetime is
\be
\begin{pmatrix} 
\gamma & \sqrt{\gamma^2-1}/c \\
c\sqrt{\gamma^2-1} & \gamma 
\end{pmatrix}.
\ee
For convenience, we choose the units of time and length to be $T$ and $a$,
respectively. In these new units, the elementary Lorentz
matrix becomes an integer matrix, reading
\be
\cal{L} = 
\begin{pmatrix} 
\gamma & 1 \\
\gamma^2-1 & \gamma 
\end{pmatrix}.
\ee
$\cal{L}$ generates a cyclic group $\tilde{\cal{L}}=\left\{\cal{L}^{\rm n} |\, {\rm n}\in\mathbb{Z}\right\}$,
dubbed the discrete Lorentz group, while the discrete Poincar\'{e} group is the direct
product of $\tilde{\cal{L}}$ and spacetime translations. The quantum theory of Poincar\'{e} crystal
is exactly the unitary representation of the discrete Poincar\'{e} group.

We consider a Poincar\'{e} crystal with periodic boundary condition,
whose Hamiltonian in momentum space reads
\be
\hat{H}_{\rm PC}=\sum_{k}E_{k}\hat{c}^\dag_{k}\hat{c}_{k}. 
\label{eq:modelPoin}
\ee
The Lorentz symmetry requires that the dispersion relation must 
be invariant under $\cal{L}$. This is a strong restriction. The method of finding 
Lorentz-invariant dispersion relations was discussed in Ref.~[\onlinecite{Wang2021JPA}]. 
An example exists as $\gamma=2$ and $L=33$ (the Lorentz
symmetry requires $L$, i.e., the lattice size, to take some specific integers). 
The corresponding dispersion relation is displayed in Table.~\ref{tab:Ek}.
Its corresponding tunneling matrix necessarily involves long-range terms.

We apply the QQFT scheme to quantum simulation of this model.
Since $L=33$ is not an integer power of $2$, for experimental realization, we perform numerical
decomposition of the Fourier matrix~\cite{Qiu2020npjQI} for local
implementation of the QQFT. In order to  benchmark the performance of our approach, we investigate single-particle quantum walks, which can be measured using quantum microscopes in cold atom experiments~\cite{2013_Bloch_Nature,2015_Greiner_Science}. 
We consider quantum walk of a single-particle initialized at site $n=0$. 
The time ($t$) evolution of its wavefunction is denoted as  $\psi_n (t)$. 
The spacetime crystallization is described by $\psi_n (t)$  at discrete times $t= s T$, with $s$ an integer and $T$ a time period determined by the Lorentz invariant energy dispersion~\cite{Wang2021JPA}. 
The symmetric properties are captured by the matrix ${\cal G}_{ns} \equiv \psi_n (s T) $. 
The discrete Lorentz symmetry implies~\cite{Wang2021JPA}, 
 \be 
 \textstyle 
{\cal G}_{ns} = {\cal G}_{n's'}, 
\label{eq:Gsymmetry} 
\ee 
with 
\be
\textstyle 
\left(\begin{array}{c}
s' \\ n' \end{array} \right) = 
{\cal{L}} 
\left(\begin{array}{c}
s \\ n \end{array} \right) \ ({\rm{mod}} \, {L}).
\label{eq:diLore}
\ee
Here, $\gamma\geq 2$ takes integer values, 
and the coordinates $n$ and $s$ are defined by modulo $L$. 
On the spacetime lattice $\left\{\left(s,n\right)\right\}$ with $0 \leq s,n  \leq L-1$,
the transformation connects one site to another. The $L\times L$ sites
are then partitioned into a few equivalence classes in each of which
the wavefunction ${\cal G}_{ns}$ must have the same value. 
As shown in Fig.~\ref{fig:PC}(a), ${\cal G}$ displays a periodic collapse and revival pattern, which respects the Lorentz symmetry (Eq.~\eqref{eq:Gsymmetry}). It has been established by one of the authors that the nontrivial quantum revival dynamics is a consequence of the Lorentz symmetry~\cite{Wang2021JPA}.

Considering the Li atom experiment setup described above, the total evolution time to simulate the Poincar\'{e} dynamics in Fig.~\ref{fig:PC} with system size $L=33$ is estimated to be $2.7$ seconds.   
This in principle can be improved by analytically decomposing  the QQFT according to $33 =3 \times 11$. 

\subsection{Robustness of simulations against noise}

To quantify the robustness against imperfections potentially existent in experiments,  
 we add Gaussian noise to the Hamiltonian sequence (Eq.~\eqref{eq:HamSeq}), 
 replacing  $H_{\rm p} ^{[s]}$ by $(1+\delta_s )H_{\rm p} ^{[s]} $, with $\delta_s$ a random variable drawn from Gaussian distribution characterized by the standard deviation $\sigma$. 
 The consequent effects on the quantum spatiotemporal dynamics are shown in Fig.~\ref{fig:PC}. 
 With increasing the noise strength, we find that the discrete Poincar\'{e} symmetry is gradually broken. The symmetric revival dynamics is evident even at a noise level of $\sigma = 5 \times 10^{-3}$. 

We provide a quantitative analysis of the noise-induced Poincar\'{e} symmetry breaking.
If a particle is located at the site $n_1$ at the initial time $t=0$, we then use
$\psi_{n_1+n}(t)$ to denote its wave function at the time $t=sT$ and the site $n_1+n$.
Then, $P_{n_1} (s,n) = \left| {\psi}_{n_1+n}(t)\right|^2$ denotes the
probability of a particle hopping from the initial site $n_1$ to the site $n_1+n$
at the time $sT$. For the conservation of probability, $\sum_{n} P_{n_1} (s, n) \equiv 1$ must hold for arbitrary $n_1$ and $s$. The translation symmetry guarantees that $P_{n_1} (s, n)$ is independent of $n_1$. More important, the Lorentz symmetry guarantees $P(s,n)=P(s',n') $ for $(s,n)$ and $(s',n')$ satisfying Eq.~\eqref{eq:diLore}.
The equivalence relation~\eqref{eq:diLore} partitions all the lattice
sites $\{(s,n)\}$ into several equivalence classes, denoted as
$C_\alpha$ with $\cup_\alpha \displaystyle C_\alpha = \{(s,n)\}$.
We use $M_\alpha$ to denote the number of sites in the class $C_\alpha$, and then have $\sum_\alpha M_\alpha=L^2$.
If the Lorentz symmetry is preserved, $P_{n_1} (s,n)$ should
be the same for those $(s,n)$ in the same equivalence class.

\begin{figure}[htp]
\centering
\includegraphics[width=.45\textwidth]{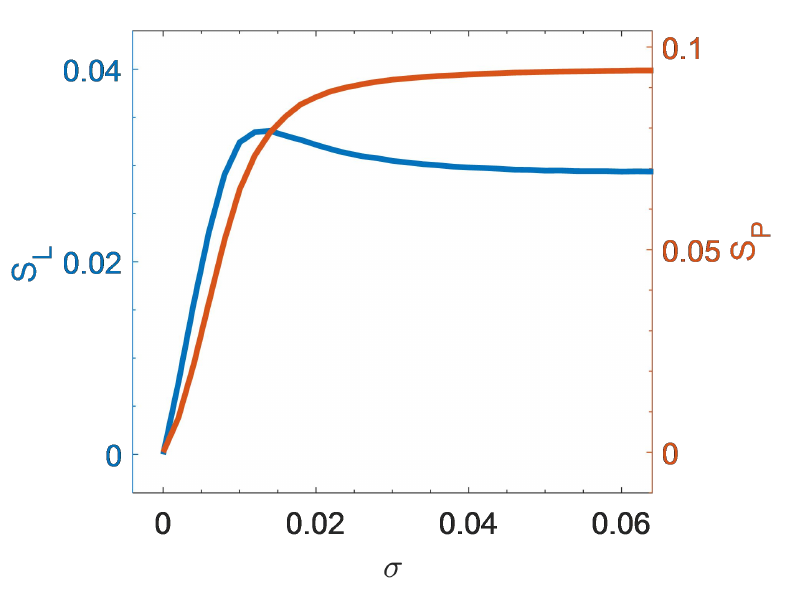}
\caption{QQFT based quantum engineering of Poincar\'{e} crystal. 
The noise strength dependence of the standard deviation $S_{\rm L}$ and $S_{\rm P}$, which are averaged over $10^2$ random noise configurations. The system size is $L=33$ and the period $T$ is set to be our time unit.
}
\label{fig:SL_SP}
\end{figure}

In order to investigate the influence of the noise to the symmetries, we define two quantities $S_{\rm L}$ and $S_{\rm P}$, which quantify the symmetry breaking in the presence of noise. We use the averaged standard deviation $S_{\rm L}=\sqrt{\frac{1}{L^2}\sum_\alpha M_\alpha S^2_\alpha}$ to quantify the Lorentz symmetry breaking, where
\bea
S^2_\alpha &=& \frac{1}{L}\sum_{n_1}\left[\frac{1}{M_\alpha}\sum_{(s,n)\in C_\alpha}\left(P_{n_1} (s,n)-\overline{P}_\alpha\right)^2\right],\nn\\
\overline{P}_\alpha &=& \frac{1}{L}\sum_{n_1}\left[\frac{1}{M_\alpha}\sum_{(s,n)\in C_\alpha}P_{n_1} (s,n)\right]. 
\eea
We thus have
\be
S_{\rm L}=\sqrt{\frac{1}{L^3}\sum_\alpha\sum_{n_1}\sum_{(s,n)\in C_\alpha}\left(P_{n_1} (s,n)-\overline{P}_\alpha\right)^2},
\ee
and $S_{\rm L}>0$ indicates that the Lorentz symmetry is broken. Similarly, we further define 
\be
S_{\rm P}=\sqrt{\frac{1}{L^3}\sum_{s,n,n_1}\left(P_{n_1} (s,n)|_{\sigma\geqslant0}-P_{n_1} (s,n)|_{\sigma=0}\right)^2},
\ee
which quantifies not only the Lorentz symmetry breaking but also the translation symmetry breaking. 
Here, $P_{n_1} (s,n)|_{\sigma}$ denotes the probability function for the noise strength being $\sigma$. We study the influence of white noise to the symmetry breaking.
Fig.~\ref{fig:SL_SP} displays $S_{\rm L}$ and $S_{\rm P}$ as a function of $\sigma$. We see that both $S_{\rm L}$ and $S_{\rm P}$ increase monotonically with $\sigma$, and reach their saturation values at $\sigma\sim 0.06$. This indicates that both the Lorentz and translation symmetries are gradually broken with increasing noise strength.

The actual noise in the DMD experiment is expected to largely depend
on technical details---it may contain both white and colored noise channels. 
By numerical simulation, we confirm the Poincar\'{e} crystal is even more
robust considering colored noise, as compared to the case of white noise shown in Fig.~\ref{fig:PC}.

Like the white noise, the colored noise is also added to the Hamiltonian sequence by replacing
$\hat H_p^{[s]}$ by $\hat H_p^{[s]}\left(1+\delta_s\right)$. The colored noise
we consider in this paper is the exponential-correlation noise, which reduces to the white noise in the limiting case. The exponential-correlation noise can be obtained by using the iterative relation~\cite{Kasdin95}
\be \label{eq:expcolornoise}
\delta_{s+1} = e^{-1/\tau_c} \delta_s +  \sqrt{1-e^{-2/\tau_c} } \sigma W_s,
\ee
where $W_s$ is a white noise with standard normal distribution, $\tau_c$ is the correlation time, and $\sigma$ is the standard deviation (noise strength). In practice, we choose $\delta_s=0$ initially, and then use Eq.~\eqref{eq:expcolornoise} to iteratively generate a time series. To remove the effect of the initial choice, we wait for $N_0$ ($N_0 \gg \tau_c$) steps before adding $\delta_s$ to the Hamiltonian sequence. The correlation function of a noise generated in this way is
\be\label{eq:colorcorr}
\langle \delta_{s} \delta_{s'} \rangle = \sigma^2 e^{-\left|s-s'\right|/\tau_c}.
\ee
The correlation decays exponentially with the time difference, and the equal-time correlation is $\sigma^2$. Furthermore, the spectral density of this noise is known to be $\sigma^2 \left(2/\tau_c\right)/\left(\omega^2+ 1/\tau_c^2\right)$, which is a Lorentzian function of the frequency $\omega$. In the limit $\tau_c\to 0$, the exponential-correlation noise reduces to a white noise of strength $\sigma$. 

\begin{figure}[htp]
\centering
\includegraphics[width=.49\textwidth]{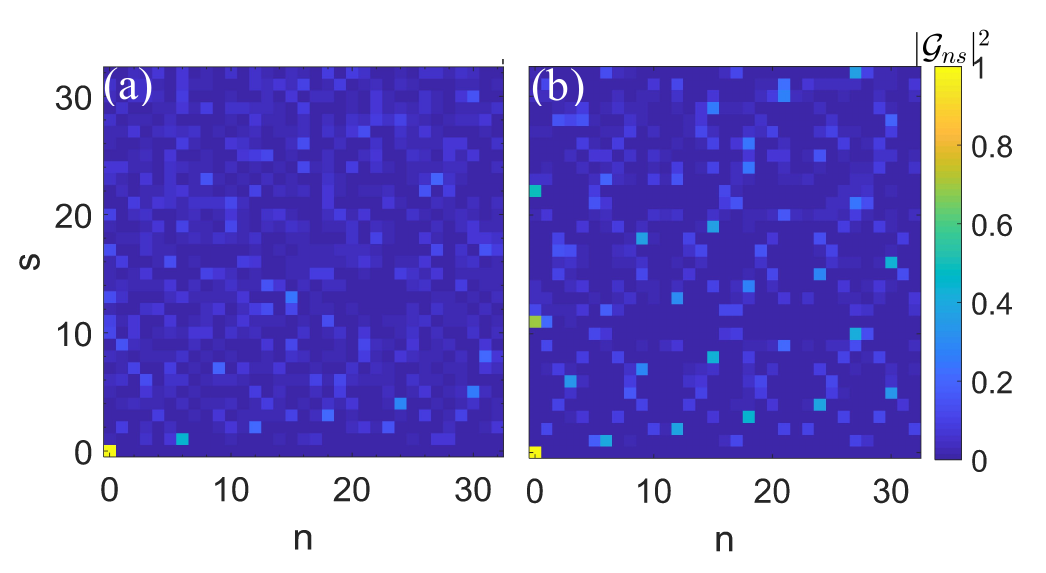}
\caption{Comparison of the effects of different noises on the single-particle wave function in a Poincar\'{e} crystal. The color indexes $| {\cal G} _{ns}|^{2}$, the probability of finding the particle at the site $n$ and the time $sT$. The strength of the noises is fixed to $\sigma=2\times10^{-2}$.  Panel (a) is for the white noise, and panel (b) is for the exponential-correlation noise with $\tau_{c}=300$. }
\label{fig:PCW}
\end{figure}

Our QQFT-based simulations are robust against reasonable colored noises. In the simulation of Poincar\'{e} crystals, the single-particle wave function preserves the full discrete-Poincar\'{e} symmetry as $\sigma=0$, but gradually loses the  symmetry as $\sigma$ (noise strength) increases. This can be seen from the melting of the spacetime-lattice structure in the wave function as $\sigma$ increases. For a given noise strength, we find that the colored noise has less effect on the wave function than the white noise.
In Fig.~\ref{fig:PCW}, we display $\left| {\cal G}_{ns}\right|^2$, where ${\cal G}_{ns} = \psi_n (s T)$ denotes the wave function at the site $n$ and the time $t=sT$ for a particle initially located at site $0$, and different panels are for the white and colored noises, respectively. While the white noise at $\sigma=2\times 10^{-2}$ destroys the Poincar\'{e} symmetry of the wave function (see Fig.~\ref{fig:PCW}(a)), the colored noise at the same strength has little effect on the wave function in which the spacetime-lattice structure and the revival and collapse pattern is clearly seen (see Fig.~\ref{fig:PCW}(b)).

\begin{figure*}[htp]
\centering
\includegraphics[width=.85\textwidth]{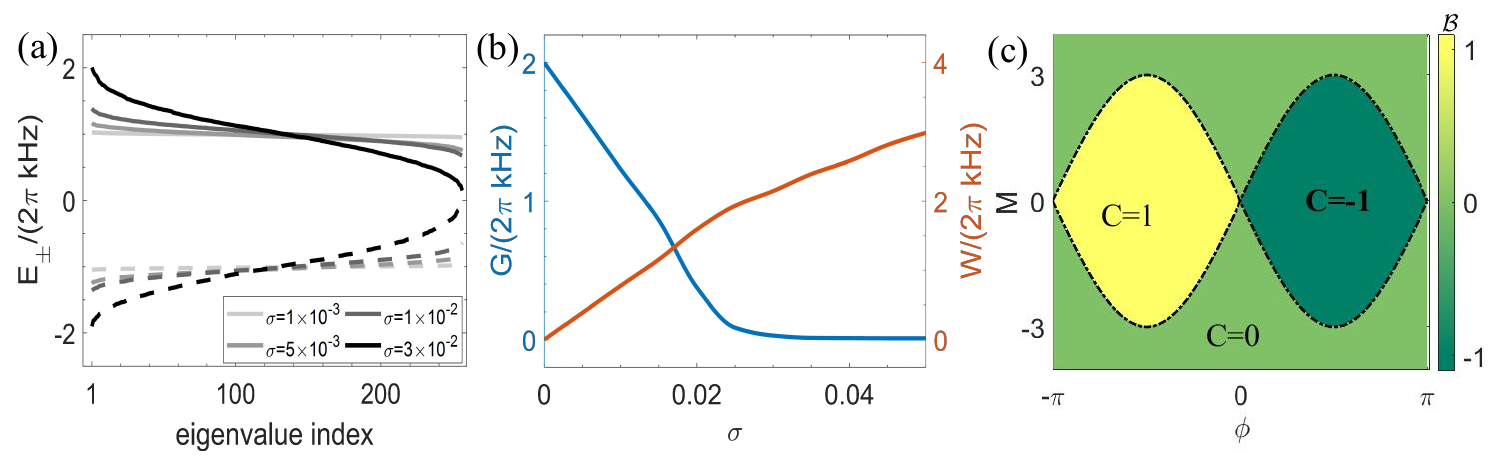}
\caption{
QQFT based quantum simulation of the flat-band Haldane model. 
(a) Upper (solid) and lower (dash-dotted) energy bands for different strength of noise. $E_+$ and $E_-$ denote the eigenenergies in the upper and lower bands, respectively.
(b) Noise strength ($\sigma$) dependence of the band gap $G$ and the band width $W$, which are averaged over $10^2$ random noise configurations.  
In (a) and (b), we set the Hamiltonian parameters $M=0$ and $\phi = -\pi/2$. 
(c) Phase diagram determined by the Bott index $\mathcal{B}$ (colormap) and the Chern number $\mathcal{C}$ (black dash-dotted lines) at the noise strength $\sigma = 3\times 10^{-2}$. We divide the Brillouin zone into a $16\times 16$ grid. 
The magnitude of ${\bf d}$ is fixed to $|{\bf d}| =2\pi$ kHz, and the evolution time is $T = 1/(2\pi)$ ms, as accessible to  cold atom experiments.}
\label{fig:CI}
\end{figure*}

\section{Quantum simulation of flat Chern bands}
\label{sec:chern}

We also apply the QQFT protocol to quantum simulation of topological flatbands. 
In the last decade, there have been great efforts on searching for 2D topological flatbands~\cite{2011_Sun_PRL,2011_Bernevig_PRX,2011_Neupert_PRL,2011_Sheng_NC,2011_Wen_PRL,2011_Wang_PRL,2012_Sondhi_PRB,2014_Roy_PRB,2015_Claassen_PRL} in both cold atom~\cite{2011_Goldman_PRA,2012_Yao_PRL,2013_Cooper_PRL,2015_Cooper_PRL,2016_Goldman_NatPhys,2021_Goldman_SPhys} and solid state systems~\cite{2020_Berghholtz_PRL,2020_Ledwith_PRR,2020_Senthil_PRR,2021_Yacoby_Nature,2021_Bergholtz_PRL}, 
as it hosts a broad range of exotic topological physics. 
The band flatness would effectively  promote strong many-body correlations, and support anomalous Landau-level physics such as 
fractional Chern insulators with repulsive interaction~\cite{2011_Bernevig_PRX,2011_Neupert_PRL,2011_Sheng_NC,2011_Wen_PRL,2011_Wang_PRL}, or high T$_{\rm c}$ topological superconductors~\cite{2015_Torma_NC,2016_Volovik,2020_Aoki,2020_Hofmann_PRB,2020_Morais_PRB} with attraction. 
It also provides an ideal platform for pure topological quantum dynamics governed by geometrical Berry curvature effects only, with non-topological dynamical response vanishing---the group velocity is zero. However, it has been proved that 2D Chern bands with complete flatness does not exist for local Hamiltonian models~\cite{2014_Chen_JPA}, causing a grand challenge to probe topological flatband physics in experiments. 

Here, we develop a scheme for engineering a completely  flat Chern band using QQFT.  
We consider a  two-band Chern insulator model in two dimensions with  
Hamiltonian 
$\hat H_{\rm CI}=
\sum_{\bf k,\alpha,\beta}\hat c^\dag_{\bf k \alpha}
 \mathcal{H}\left({\bf k}\right)_{\alpha,\beta} \hat c_{\bf k \beta}$.
In general, the Hermitian matrix $\mathcal{H}\left({\bf k}\right)$ 
can be written in terms of Pauli matrices as 
\be
\textstyle 
\mathcal{H}\left({\bf k}\right)=d_0({\bf k}) \sigma_0+{\bf d}({\bf k})\cdot{\boldsymbol \sigma}, 
\label{eq:CI}
\ee
with $\sigma_0$ is the identity, 
${\boldsymbol \sigma}=(\sigma_1,\sigma_2,\sigma_3)$
 the Pauli matrices, 
 and ${\bf d}=(d_1,d_2,d_3) $ a three-component vector. 
 To engineer flat bands, we  choose $d_0({\bf k}) = 0$, and set 
\be 
\textstyle  {\bf d}({\bf k}) \cdot {\bf d}  ({\bf k}) = {\rm const}. 
 \label{eq:dnorm} 
\ee 
With conventional quantum simulation schemes, this condition is almost impossible to reach as it requires delicate long-range couplings. 
In contrast, with our QQFT approach, engineering the momentum space Hamiltonian ${\bf d}({\bf k})$ is reached by  programming the local potential in the physical system, which is accessible to cold atom experiments~\cite{2016_Weiss_Science,browaeys2020many,2021_Sengstock_Nature,Qiu2020npjQI}. 
The real-space lattice in the experimental realization corresponds to discretizing the Brillouin zone of the model in our scheme. 
We impose a square lattice grid of size $L\times L$ on the Brillouin zone, for which the required 2D QQFT in experimental realization is separable simply containing two successive 1D QQFTs, as explained
in Sec.~\ref{sec:genqqft}.

The realized  Chern band with our approach is completely flat by design. 
To investigate its robustness against experimental imperfections,  
we add Gaussian noise to the Hamiltonian sequence as in the previous example, 
 and examine the band flatness and the topological property. 
For a concrete demonstration, we consider the Haldane honeycomb lattice 
model~\cite{Haldane1988PRL}, which consists of two sublattices
$A$ and $B$. The Hamiltonian has a form in Eq.~\eqref{eq:CI}, with 
$d_1({\bf k})= t_1\left[1+\cos\left({\bf k}\cdot{\bf g}_2\right)+\cos\left({\bf k}\cdot{\bf g}_3\right)\right]$,
$d_2({\bf k})=t_1\left[\sin\left({\bf k}\cdot{\bf g}_2\right)-\sin\left({\bf k}\cdot{\bf g}_3\right)\right]$,
and $d_3({\bf k})=M-2t_2\sin\phi\sum^3_{j=1}\sin\left({\bf k}\cdot{\bf g}_j\right)$.
Here, the primitive lattice vectors ${\bf g}_1={\bf e}_2-{\bf e}_3$, ${\bf g}_2={\bf e}_3-{\bf e}_1$, and ${\bf g}_3={\bf e}_1-{\bf e}_2$, where $\bf{e}_1$, $\bf{e}_2$, and $\bf{e}_3$ are the three unit vectors pointing from one $B$ lattice site to its three neighboring $A$ sites on the honeycomb lattice.
We choose $t_2/t_1 = 1/\sqrt{3}$ here. 
In our flatband engineering protocol, we  normalize the ${\bf d}$ vector, i.e., 
${\bf d} \to {\bf d}/|{\bf d}|$.

The energy spectrum and the topological property are obtained by decomposing the dynamical evolution operator  
$\mathcal{U}_{\rm CI} = e^{-iH_{\rm CI} T}$, i.e., 
$\mathcal{U}_{\rm CI}\ket{{n, \alpha}} =\exp\left(-\im T E_{n, \alpha}\right)\ket{n, \alpha}$, 
with $\alpha=+$($-$) and $n\in [1, L^2]$ indexing the upper (lower) band, and  the eigenstates within each band, respectively. 
To demonstrate the band flatness, 
we calculate the band gap $G=\min \{E_{{n},+}\}-\max \{E_{{n},-}\}$ and the band width $W =\max \{E_{{n},-}\}-\min \{E_{{n},-}\}$ in presence of noise.  
We find that the band gap (band width) decreases (increases) with the strength of the Gaussian white noise, 
and confirm that the QQFT protocol is reasonably robust---the band flatness ratio $W/G$ remains below $10\%$ even at $\sigma = 2.5\times10^{-3}$ (Fig.~\ref{fig:CI}(a,b)). 
The required noise level is below $10^{-2}$, and is close to the quantum  control precision demonstrated in experiments~\cite{lukin2019probing,2021_Zhou_PRAL}, and is anticipated to be reachable in near terms with the recently developed calibration protocol~\cite{Qiu2020npjQI}. 
For the topological property, we compute Bott index (${\cal B}$)~\cite{Hastings2010JMP, Loring_2010EPL, Hastings2011AoP}, 
which is a topological invariant irrespective of the noise induced  translation symmetry breaking (see Fig.~\ref{fig:CI} (c)). The Bott index of our topological flatband is equal to the Chern number 
$\mathcal{C}$~\cite{Haldane1988PRL} in the clean limit. In presence of noise, the Bott index remains quantized and unaffected as long as the band gap remains open. 

For the experimental realization with Li atoms, the QQFT based quantum engineering of the topological Chern band takes about 
$100$ ms to simulate  lattice with size $16\times 16$. 
We show a topological flat band can be achieved with bandgap and bandwidth  reaching $2\pi \times 2$~kHz and $2\pi \times 0.2$~kHz, assuming a noise level $\sigma=2.5 \times 10^{-3}$ (Fig.~\ref{fig:CI}). We expect this realization to be accessible to cold atom experiments. 

\begin{figure}[htp]
\centering
\includegraphics[width=.49\textwidth]{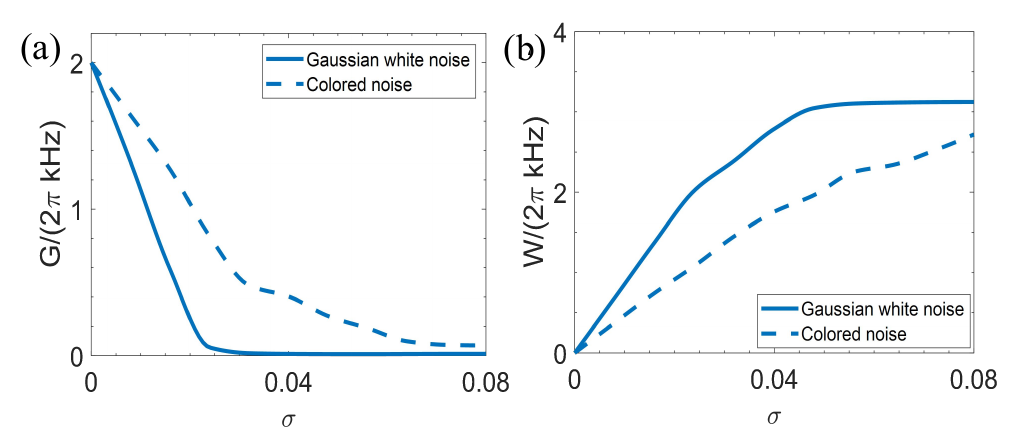}
\caption{Comparison of the effects of different noises on the energy bands of a flat-band Haldane model. Panels (a) and (b) display the band gap and band width, respectively, as a function of noise strength, which are averaged over $200$ random noise configurations. Lines of different types are for the white noise (solid) and exponential-correlation noise with $\tau_c=300$ (dash-dotted), respectively. In the simulation, we consider a $16\times 16$ lattice. The Hamiltonian parameters are chosen to $M=4$ and $\phi = -\pi/2$. The magnitude of ${\bf d}$ is fixed to $|{\bf d}| =2\pi$ kHz, and the evolution time is $T = 1/(2\pi)$ ms, as accessible to  cold atom experiments.}
\label{fig:CIW}
\end{figure}

The simulation of flat-band Chern insulators is also
robust against the colored noise that has an exponential correlation (see Eq.~\eqref{eq:colorcorr}).
For the colored noise, the energy gap also gradually closes with increasing $\sigma$.
But once if the gap is open, the noises do not change the Bott index, which always equals the ground-state Chern number of the Haldane model. In Fig.~\ref{fig:CIW}, we compare the effects of different noises on the energy gap and bandwidth. For a given $\sigma$, the colored noise has less effect on the energy bands, in comparison with the white noise. This is seen from the slower dropping (increasing) of the band gap (width) in the presence of colored noise, as $\sigma$ increases. 

For both the Poincar\'{e} crystal and flat-band Haldane model, the colored noise has less effect on our QQFT-based simulations, in comparison with the white noise. A rough explanation in the limiting case is given below. If we consider the dc-noise limit ($\tau_c\to\infty$), $\delta_s$ then becomes a $s$-independent constant. The dc-noise modifies both the QQFT ($\hat{V}$) and its Hermitian adjoint ($\hat{V}^\dag$) by replacing the Hamiltonian $\hat{H}_p$ by $\hat{H}_p\left(1+\delta_s\right)$. If $\delta_s$ is a constant, then such a replacement keeps $\hat{V} \hat{V}^\dag = 1$. Because the evolution operator is $\hat U = \hat{V} e^{-i T \hat{H}_D}\hat{V}^\dag$, the modification to $\hat{V}$ has no influence on the spectrum of $\hat U$. This explains why a dc-noise has less effect on the energy bands than a white noise.

\section{Conclusion and outlook}
\label{sec:con}

We propose a QQFT based Hamiltonian  engineering scheme, which allows  flexible quantum simulations of translationally invariant Hamiltonian models. The Hamiltonian control sequence to implement QQFT is analytically constructed, considering cold atoms confined in a programmable optical potential. We demonstrate its capability  by investigating quantum simulations on the Poincar\'{e} spacetime crystal and topological flatband, both of which are extremely challenging to realize with conventional approaches for the required delicate long range couplings. Its robustness against imperfections potentially existent in optical lattice experiments has been confirmed by numerical simulations. 

\begin{acknowledgments}
{\it Acknowledgement.---} 
We appreciate helpful discussion with Andreas Hemmerich, Lei Shi, Xiaoting Wang, and Hanning Dai. 
The work is supported by National Program on Key Basic Research Project of China (Grant No. 2021YFA1400900), National Natural Science Foundation of China (Grants No. 11835011, 11774315, 12104098, 11934002), Shanghai Municipal Science and Technology Major Project (Grant No. 2019SHZDZX01), Shanghai Science Foundation (Grants No.21QA1400500, 19ZR1471500), 
and the Junior Associates program of the Abdus Salam International Center for Theoretical Physics. 
\end{acknowledgments}

\bibliography{references}

\end{document}